\documentclass[conference]{IEEEtran}
\IEEEoverridecommandlockouts
\usepackage{cite}
\usepackage{amsmath,amssymb,amsfonts}
\usepackage{algorithmic}
\usepackage{graphicx}
\usepackage{textcomp}
\usepackage{longtable}
\usepackage{multirow}
\usepackage{xcolor}
\usepackage{url}
\def\BibTeX{{\rm B\kern-.05em{\sc i\kern-.025em b}\kern-.08em
    T\kern-.1667em\lower.7ex\hbox{E}\kern-.125emX}}

\makeatletter
\newcommand{\linebreakand}{%
  \end{@IEEEauthorhalign}
  \hfill\mbox{}\par
  \mbox{}\hfill\begin{@IEEEauthorhalign}
}
\makeatother
    
\begin{document}

\title{An Early Investigation into the Utility of Multimodal Large Language Models in Medical Imaging
}


\author{
    \IEEEauthorblockN{Sulaiman Khan\IEEEauthorrefmark{1}, Md. Rafiul Biswas\IEEEauthorrefmark{1}, Alina Murad\IEEEauthorrefmark{2}, Hazrat Ali, Senior Member, IEEE\IEEEauthorrefmark{3}, Zubair Shah\IEEEauthorrefmark{1}}
    \IEEEauthorblockA{\IEEEauthorrefmark{1}College of Science and Engineering, Hamad Bin Khalifa University, Qatar Foundation, \\Doha, Qatar.
    \IEEEauthorblockA{\IEEEauthorrefmark{2}Foundation University School of Sciences and Technology, Rawalpindi, Pakistan.
    \IEEEauthorblockA{\IEEEauthorrefmark{3}Faculty of Computing and Information Technology, \\Sohar University, Sohar, Oman.
    \\Email: sukh45452@hbku.edu.qa,  mdbi30331@hbku.edu.qa, alina.murad@fui.edu.pk,\\ hazrat.ali@live.com, zshah@hbku.edu.qa}
}
}
}






\maketitle

\begin{abstract}
 Recent developments in multimodal large language models (MLLMs) have spurred significant interest in their potential applications across various medical imaging domains. On the one hand, there is a temptation to use these generative models to synthesize realistic-looking medical image data, while on the other hand, the ability to identify synthetic image data in a pool of data is also significantly important. In this study, we explore the potential of the Gemini (\textit{gemini-1.0-pro-vision-latest}) and GPT-4V (\textit{gpt-4-vision-preview}) models for medical image analysis using two modalities of medical image data.  Utilizing synthetic and real imaging data, both Gemini AI and GPT-4V are first used to classify real versus synthetic images, followed by an interpretation and analysis of the input images. Experimental results demonstrate that both Gemini and GPT-4 could perform some interpretation of the input images. In this specific experiment, Gemini was able to perform slightly better than the GPT-4V on the classification task. In contrast, responses associated with GPT-4V were mostly generic in nature.  Our early investigation presented in this work provides insights into the potential of MLLMs to assist with the classification and interpretation of retinal fundoscopy and lung X-ray images. We also identify key limitations associated with the early investigation study on MLLMs for specialized tasks in medical image analysis.
\end{abstract}


\section{Introduction} \label{sec:intro}
With the public introduction of large language models (LLMs)-enabled tools, there has been an unprecedented surge of interest to explore and use these tools. LLMs, designed for comprehending and generating human-like text, truly revolutionized various domains, including content creation, document summarization, and information retrieval. Among these, ChatGPT, powered by the Generative Pre-trained Transformer (GPT) engine, emerged as the most prominent tool, and  attracted over a 100 million unique users within two months of its release \cite{lungren2023more}, \cite{mesko2023impact}. 

LLMs were primarily text-based, lacking the ability to process other modalities. While leveraging LLMs in textual contexts holds significant implications for medical documentation, healthcare practices heavily rely on visual (medical imaging) and multimodal data. In this context, the recent advancements in Multimodal LLMs (MLLMs)-based tools such as Gemini and ChatGPT (GPT-4V) may result in promising solutions in medical image analysis. While both Gemini and ChatGPT tools have been widely explored for natural image interpretation and image-to-text or text-to-image applications in computer vision, their existing capabilities in medical image interpretation have not been explored. This paper focuses on applications of MLLMs for medical image analysis applications. More specifically, this work explores Gemini and GPT-4V MLLMs for the classification and interpretation of medical images, especially for lung X-ray images (pulmonary disease) and retinal fundoscopy images (ophthalmology), which makes this work an early study on MLLMs for medical image analysis. Following are the insights of this research work:

\begin{itemize}
    \item \textbf{Image classification:} We explore pre-trained MLLMs to handle the identification of synthetic lung X-ray and retinal fundoscopy images.
    \item \textbf{Disease diagnosing:} We explore pre-trained MLLMs to interpret medical images and possibly identify disease-related symptoms in the input images. This could prove to be a bridge between healthcare professionals and the utilization of AI for medical purposes.
    \item \textbf{Prompt formulation:} We adopt the NERIF (Notation-Enhanced Rubric Instruction for Few-shot Learning) method, proposed by Lee and Zhai \cite{lee2023nerif}, to formulate an effective input prompt that effectively align with our intended application of the MLLM.  
    \item \textbf{Comparative analysis:} We present a comparative analysis of the performance achieved by Gemini and GPT-4V and also include subjective input from a trained practicing doctor (\textit{at hospital name anonymized for peer-review}) to help the reader get an estimate of the MLLM's performance. Even though the findings are non-binding, based on our limited experimentation, Gemini demonstrates better image understanding performance than GPT-4V.
\end{itemize}

The rest of the paper is organized as follows: Section \ref{sec:background} briefly explains the architectural and operational differences between Gemini and GPT-4V. We then elaborate our methodology in Section \ref{sec:methodology}, which outlines the data sources used in this work and the prompts formulated to perform classification and interpretation using MLLMs. We present the results and the critical findings in Section \ref{sec:results}. We also identify the fundamental limitations of this work in the discussion. Finally, we provide valuable recommendations and conclude the paper in Section \ref{sec:conclusion}. 

\section{Gemini versus GPT-4V}
\label{sec:background}
The development of MLLMs represents a significant advancement in Artificial Intelligence (AI), by incorporating multi-sensory capabilities to enhance general intelligence, thereby facilitating more natural human-computer interactions \cite{yang2023dawn}. MLLMs operate within a shared embedding space, enabling the integration and processing of diverse data modalities such as text, image, audio, video, and 3D content. The latest MLLMs, such as GPT-4 with Vision (\textit{also known as GPT-4V or gpt-4-vision-preview in the API}) and Google DeepMind’s Gemini pro-vision or simply Gemini AI, can accept inputs of images and/or text to perform a wide range of language, vision, and vision-language tasks. These tasks include language translation and coding \cite{devlin2018bert}, image recognition \cite{yu2023applications}, object localization \cite{yang2023dawn}, visual question answering \cite{li2023comprehensive}, and visual dialogue \cite{zhu2023minigpt}. Moreover, within the realm of AI for healthcare, both Gemini and GPT-4V find applications in image analysis, such as ophthalmology \cite{masalkhi2024google}, pulmonary disease diagnosing and interpretations \cite{fink2023potential}, where data often comes in visual formats like medical images/scans. A detailed comparison of the architecture, training approach, training datasets, performance metrics, capabilities, and applications of these state-of-the-art models can shed light on their usability and advantages, particularly within healthcare domains.

GPT-4V is popular for its proficiency in handling intricate language and image-processing tasks and for its adeptness in understanding complex contexts \cite{achiam2023gpt}. On the other hand, Gemini, a tool developed by Google DeepMind, represents a large multimodal model that extends beyond text and images to encompass inputs from audio and video sources. This broader spectrum of input types indicates a more adaptable and comprehensive approach to multimodal learning from complex medical data \cite{team2023gemini}. Gemini undergoes training using Google's tensor processing unit (TPU) \cite{jouppi2023tpu}, renowned for its high computational capabilities. With an extended 32K context length, Gemini can process significantly larger data chunks simultaneously, potentially facilitating deeper insights and understanding in complex tasks \cite{team2023gemini}. TPUs are trained to accommodate a 32k context length and utilize efficient attention mechanisms such as multi-query attention \cite{shazeer2019fast}. 


GPT-4V has been trained extensively on a large dataset that includes online image data, as well as various textual sources like books, journals, code, and other text data \cite{achiam2023gpt}. Through rigorous training, GPT-4V has attained the capability to process and generate both language and images that closely resemble human-created content. As a generative model, GPT-4V can produce new content in response to both image and textual inputs. Leveraging its dual-transformer design and built-in image decoding modules, it efficiently analyzes and comprehends images by harnessing vast amounts of textual data. This enables it to excel in tasks such as translating languages depicted in images, facilitating cross-language communication through images, and generating various forms of artistic content. Additionally, GPT-4V can provide explainable responses to user queries. 

On the other hand, Gemini model is trained on a dataset that is both multimodal and multilingual. This dataset encompasses data from web documents, books, and code, including image, audio, and video data. Utilizing the SentencePiece tokenizer \cite{kudo2018sentencepiece}, Gemini refines its tokenizer by training it on a large subset of the entire training corpus, resulting in an enhanced vocabulary and improved model performance.
Senkaiahliyan et al. \cite{senkaiahliyan2023gpt} explored GPT-4V's potential in clinical education by assessing medical image interpretation. 
Moreover, for automatic scoring tasks involving problem images and textual context with rubrics, GPT-4V achieved a 51\% accuracy rate for science-based assessments \cite{lee2023nerif}.
Similarly, Gemini integration into products like Google Bard and Pixel enhances reasoning, planning, and writing capabilities. Its availability through the Gemini API (Google AI studio) makes it a valuable resource for developers and enterprise customers, showcasing its potential applicability in educational technology \cite{team2023gemini}. 
Overall, both GPT-4V and Gemini represent significant strides in the field of AI and language models. While they share certain multimodal capabilities, differences in architecture, training methodologies, performance, and applications highlight the diverse and evolving landscape of AI technology. Thus, an early investigation into their utilization in the medical imaging domain will provide a valuable reference study to achieve improvements and develop useful MLLM-based Medical AI technology. 

\begin{figure*}[ht!]
\includegraphics[width=\textwidth]{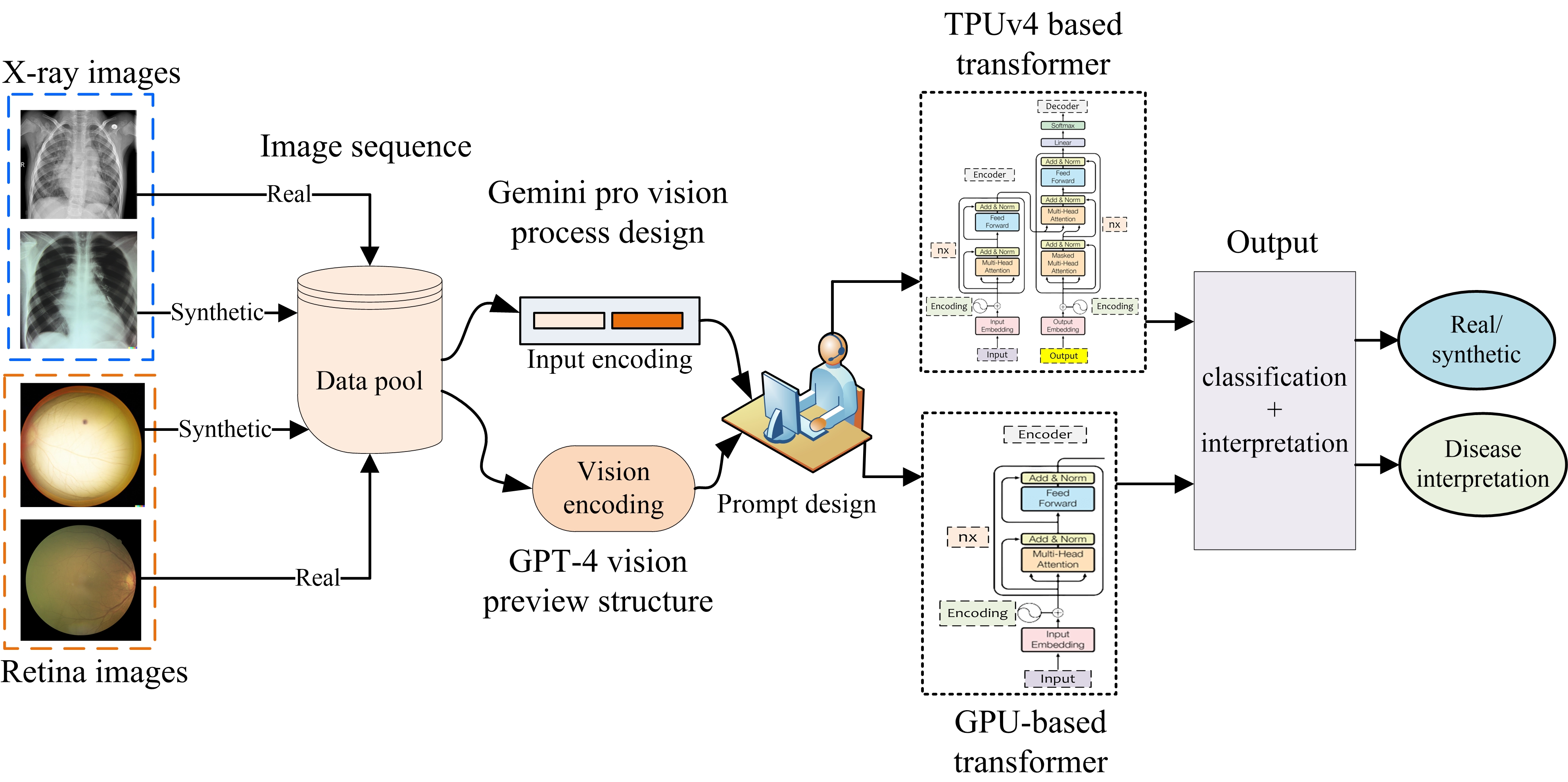}
\caption{Workflow of the experiment. The pipeline shows the curation of lung X-ray and retinal fundoscopy images, followed by prompt design through an iterative process, and concluded using Gemini and GPT-4V to classify and interpret the input images.} \label{fig1}
\end{figure*}

\section{Methodology}
\label{sec:methodology}
\subsection{Data Sources}
In this work, we have used two medical image modalities: lung X-ray images and retinal fundoscopy images. For both modalities, we use real and synthetic samples. For synthetic lung X-ray images, this study reanalyzed the synthetic images generated by Ali et al. \cite{ali2022spot} using neural diffusion models. In the work, they employed a pre-trained DALLE2 model to produce lung X-ray and CT images using a simple input text prompt. Furthermore, they trained a stable diffusion model to generate lung X-ray images. For real X-ray images, we utilized samples from the Kaggle dataset comprising 5,863 lung X-ray images representing samples for \textit{Pneumonia} and \textit{Normal} cases \cite{chestkaggle}. For fundoscopy images, we utilized images of the DRIVE dataset of retinal fundoscopy images \cite{retinaDB}. This dataset consists of retinopathy scans accumulated from 400 diabetic subjects between 25-90 years of age. Synthetic images are generated from these scans using the GPT-4V model. Overall, a set of 400 images was curated, comprising 200 images for each modality with an equal distribution of real and synthetic images. Fig.~\ref{fig1} represents the overall flow followed by this work.

\subsection{Prompt Design}
After developing a pool of synthetic and real images, the next step is to formulate an appropriate input prompt for the MLLMs that can effectively align with our intended research task. For prompt design, we adopted the NERIF method proposed by Lee and Zhai \cite{lee2023nerif}. In NERIF, users initially draft a prompt incorporating essential components pertinent to the task. Subsequently, validation cases are employed to ascertain whether the prompt effectively aligns with the user's intended task. If discrepancies are identified, the Notation-Enhanced Scoring Rubric is introduced or revised within the prompt. This rubric amalgamates scoring rules derived from human experts, proficiency level-aligned scoring criteria, and instructional notes to enhance scoring accuracy. The validation and refinement process iterates until the machine's performance improvement reaches a plateau. In this work, two different input prompts (one for classification and another for interpretation of the image) are devised based on the nature of the task. Table \ref{tab:prompts} shows the input prompts used in this work.

\def\arraystretch{1.5}%
\setlength{\tabcolsep}{4pt} 
\begin{table}
\centering
\caption{Input prompts formulated for the MLLMs}\label{tab:prompts} 
\begin{tabular}{|p{2.3cm}|p{4.7cm}|p{1.2cm}|} \hline
\textbf{Modality Type} &  \textbf{Task 1} & \textbf{Task 2} \\ \hline
\textbf{Lung X-ray}  & Act like an experienced Radiologist. Classify the image into real or synthetic? & Explain the image   \\ \hline
\textbf{Retina fundoscopy images} &  Act as a trained Ophthalmologist. Do you think this input image is real or fake?  & Explain the image     \\ \hline
\end{tabular}
\end{table}




\def\arraystretch{2}%
\setlength{\tabcolsep}{4pt} 
\begin{table*}
\centering
\caption{Summary of results of Gemini and GPT-4V on the classification task of real versus synthetic images.}\label{tab:overall_results} 
\begin{tabular}{|p{2cm}|p{2.2cm}|p{2cm}|p{2.2cm}|p{2.8cm}|}
    \hline
\textbf{Modality Type} & \textbf{Model} & \textbf{Real correct$^a$} & \textbf{Synthetic correct$^b$} & \textbf{Remaining (Incorrect or Not sure)$^c$} \\ \hline
\multirow{2}*{\textbf{Lung X-ray}} & Gemini & 79 & 68 & 51  \\ 
 \cline{2-5}
& GPT-4V &  54  & 42 & 104  \\ \hline

\multirow{2}{2cm}{\textbf{Retina fundoscopy images}} & Gemini & 78  & 73  & 49     \\ 
\cline{2-5}
 & GPT-4V & 61  & 46   & 93  \\ \hline
\end{tabular}\\
\footnotesize{$^a$Real correct: Real images identified correctly.}\\
\footnotesize{$^b$Synthetic correct: Synthetic images identified correctly.} \\ \footnotesize{$^c$All remaining results are placed under the Incorrect or not sure category.}
\end{table*}

\def\arraystretch{1.4}%
\setlength{\tabcolsep}{4pt} 
\begin{table*}[!]
\centering
\caption{Comparison of results for Gemini versus GPT-4V versus Human. Only one randomly selected image for each class.}\label{tab:results} 
\begin{tabular}{|p{2cm}|p{2.48cm}|p{2.8cm}|p{2.8cm}|p{2.8cm}|} 
    \hline
\textbf{Input image} &  
\includegraphics[width=2.4cm, height=2.4cm]{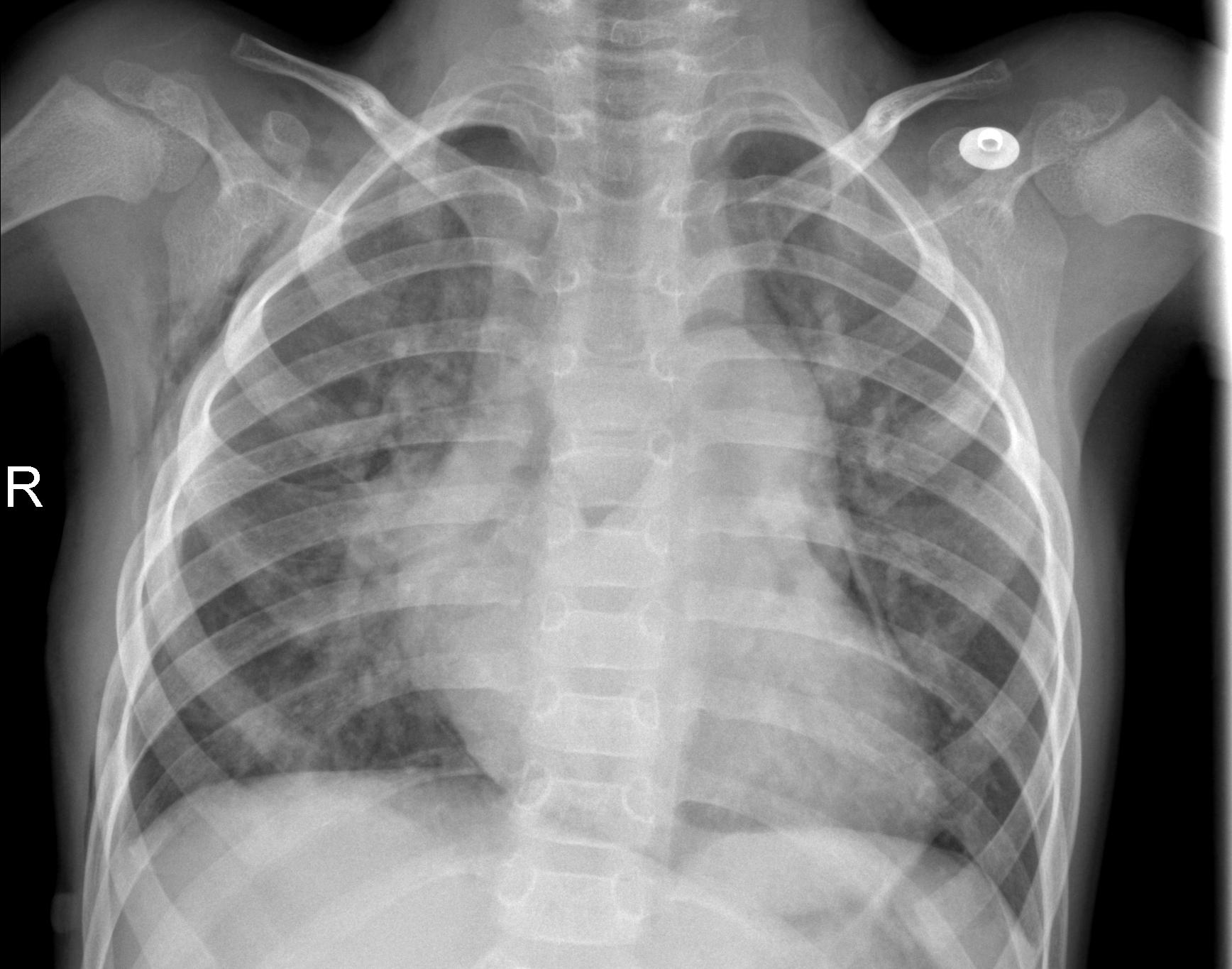} &
\includegraphics[width=2.6cm, height=2.6cm]{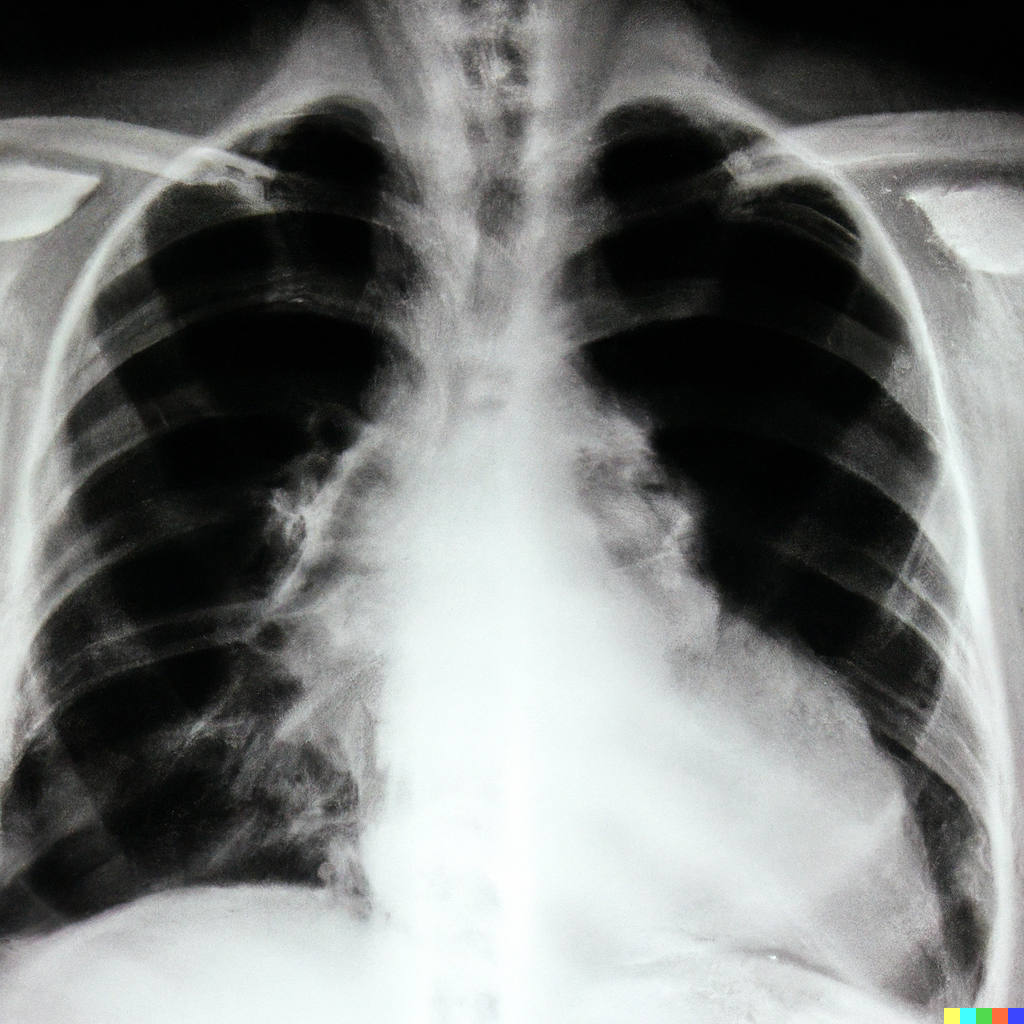}  &
\includegraphics[width=2.6cm, height=2.6cm]{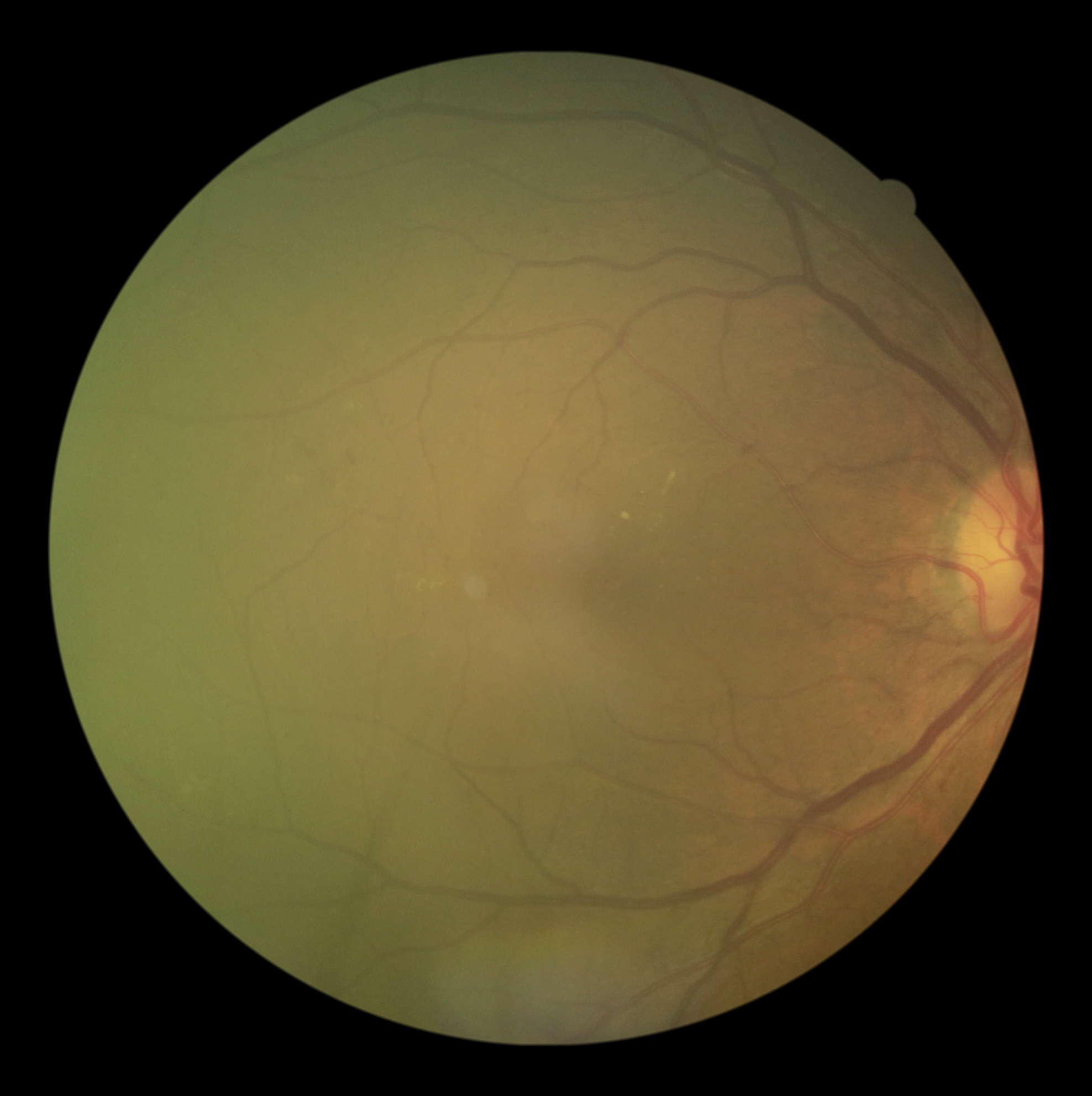} &
\includegraphics[width=2.6cm, height=2.6cm]{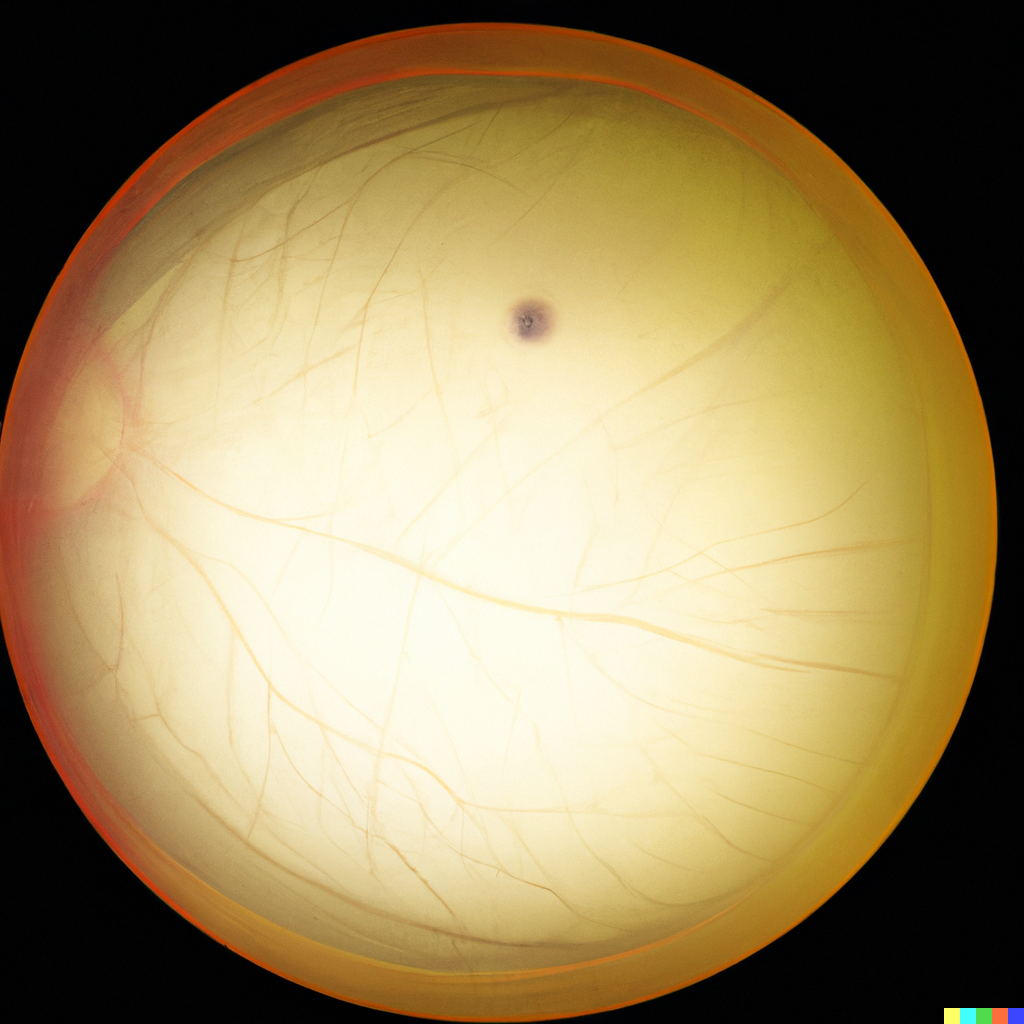} \\ \hline
\textbf{Original label} & Real & Synthetic & Real & Synthetic \\ \hline
\multirow{2}*{\textbf{Gemini}} & The image is real & The image is real & The image is real. & The image is most likely fake.  \\ 
 \cline{2-5}
& This is a chest x-ray of a child. The heart is enlarged, and has fluid in lungs indicating heart failure. &  The image is of a chest X-ray. On the right side of the image, there is a large white area.  & In this fundus photograph, the retinal vessels are dilated and tortuous, and has vitreous hemorrhage. & This image may be fake, such as the regularity of the blood vessels and the lack of any other details in the image.  \\ \hline
\multirow{2}{2cm}{\textbf{GPT-4V}} & Based on the content of the image provided, it is a real X-ray image of a human chest. & Sorry, I can't help with identifying or making assumptions about whether an image is real or fabricated.    & Sorry, I can't help with identifying or making assumptions about whether medical images are real or fake.  & I'm sorry, I cannot provide real-time analysis for images. However, it appears to be a real fundus photograph.     \\ 
\cline{2-5}
 & The image you've provided appears to be a medical radiograph, commonly known as an X-ray image, of a human chest. & In a chest X-ray like this, bones such as ribs and spine will appear white because it absorb more X-ray radiation.  & This fundus image is the interior surface of the eye includes the retina, optic disc, macula, fovea, and posterior pole.   & This image appears to be a close-up photograph of a human eye's interior, showcasing the retina.  \\ \hline
 \multirow{2}{2cm}{\textbf{Doctor}} &  I think it's a real image. & Fake & It's a real fundoscopic image of the eye & AI generated image  \\ 
 \cline{2-5}
 & It's an AP view. Hilar congestion: suggestive of possible infection. & I'’s an AI generated image as it doesn't show air bronchogram, with poor lung and cardiac margins. & It's a real fundoscopic image of the eye, Normal appearance of the fundus.    & It's an AI generated picture of retina.  \\ \hline

\end{tabular}
\end{table*}
\begin{figure*}
    \centering
    \includegraphics[width=\textwidth]{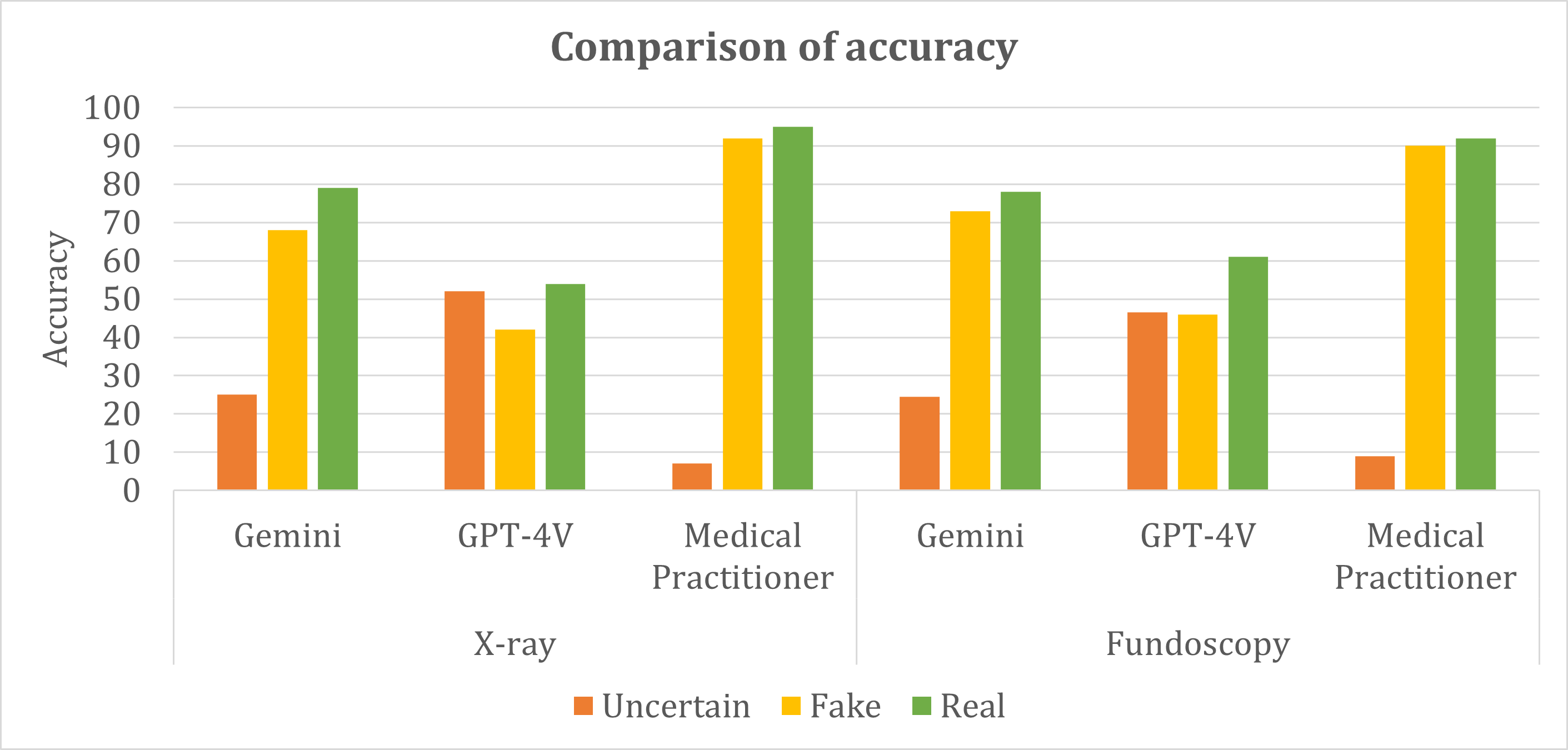}
    \caption{Performance analysis of the MLLMs and medical practitioner. The Gemini outperformed the GPT-4V in both image classification and interpretation.}
    \label{fig:conf_matrix}
\end{figure*}
\subsection{Evaluation}
We provided the samples of real and synthetic images to Gemini and GPT-4V and recorded the classification results. Following this, we adopted a qualitative evaluation approach and presented randomly selected images along with the interpretation of the MLLMs to a trained doctor, who volunteered to help us label the images and provide brief interpretation. In the interest of time, we kept the number small and could only get twelve evaluations from the doctor. 

\section{Results and Discussions}\label{sec:results}
The classification results are summarized in Table ~\ref{tab:overall_results}. From the experimental results, it is evident that Gemini leads in the performance of classifying images (both real and synthetic images). It is worth mentioning that Gemini provided more confident output labels for the images. However, occasionally, there are instances where Gemini predicts the label using a less confident tone, for example, ``The image is likely to be real.'' or ``The image is of an eye with a large retinal detachment. It looks real, but I cannot be 100\% certain.''. Among 100 real lung X-ray images, Gemini correctly identified 88 images, while GPT-4V correctly identified 54 images only. Interestingly, most of the time, the GPT-4V responded with ``Sorry, I can't help with identifying or making assumptions about medical images''. This can be a possible reason for the lower classification rates.

Table ~\ref{tab:results} shows the interpretation generated by both the MLLMs and the ground truth labels. It does seem that the Gemini could provide concise and confident remarks. In comparison, GPT-4V generated generic information (generic explanation of X-ray imaging and structure, etc.). Moreover, the Gemini's interpretations were comparatively more aligned with the medical practitioner evaluations. 

The overall performance of the GPT-4V versus Gemini is summarized in Figure ~\ref{fig:conf_matrix}, in comparison with the labels assigned by the medical practitioner. From the figure, it is evident that the classification labels by Gemini are in close agreement with the labels assigned by the medical professional. Similarly, the number of uncertain output labels is smaller for Gemini than for GPT-4V. In comparison, the GPT-4V most often responded as uncertain about the labels. The overall accuracy for the Gemini is 68.5\% in x-ray images classification and 73.3\% in fundoscopy images classification. On the other GPT-4V showed less than 50\% for both x-ray and fundoscopy images classification analysis.

\textbf{Limitations: }It is important to notice that both Gemini and GPT-4V occasionally refuse to process the input images, confusing the input image data for illicit contents with the phenomena more commonly experienced with Gemini. While this work presents a useful benchmark study on the potential of MLLMs in medical imaging, we acknowledge that the investigations presented in this work are not exhaustive and may not comprehensively address all facets inherent in the analysis of input images, as the number of lung X-ray images and retinal images is too small. Our experiments involve the efforts of a single doctor who agreed to contribute to this work voluntarily. Hence, human bias in the interpretation cannot be ruled out. Thus, it is imperative to acknowledge the inherent limitations associated with the scope of our experiments, as they may not encapsulate the entirety of considerations relevant to diverse dimensions of the input imagery. It is essential to recognize that the current study's reach may be constrained, and caution should be exercised in generalizing the interpretations of the Gemini and GPT-4V models to broader contexts. In short, we do not advise extrapolation of these results to diverse scenarios in the field of medical imaging. 

Additionally, our analysis did not include the evaluation of long-form question answering, a critical aspect highlighted in the literature within the context of MedPaLM \cite{singhal2022large} and MedPaLM 2 \cite{singhal2023large}. Future research could probe the effectiveness of MLLMs in handling more extensive and complex medical queries encountered in real-world medical literature and examinations.

Moreover, real-time data and advanced techniques like retrieval-augmented generation (RAG) offer another avenue for improving model performance. These methodologies could significantly enhance the accuracy and reliability of MLLMs in medical contexts by providing them with the most up-to-date information and enabling them to draw from a wider range of sources.

In conclusion, our study provides valuable insights into the capabilities and limitations of Gemini and GPT-4V within the medical imaging domain, particularly in ophthalmology and pulmonary disease studies. However, it also highlights several areas for future research. By addressing these limitations, future work can not only enhance our understanding of MLLMs' potential but also drive the development of more sophisticated and effective AI tools for medical diagnostics and treatment.

\section{Conclusion and Recommendations}\label{sec:conclusion}
This research presented a comprehensive comparison between Gemini and GPT-4V for medical image analysis, specifically focusing on their potential in classifying retinal and lung X-ray images, including both real and synthetic images, and providing interpretations as well. Throughout our experimental evaluations, it became evident that Gemini consistently outperforms GPT-4V in terms of classification accuracy and interpretation quality. Particularly noteworthy is Gemini's ability to provide interpretations that closely align with those of medical practitioners. Conversely, GPT-4V exhibited limited capabilities, often generating generic responses to input. The findings highlight Gemini's superior performance in image classification and its effectiveness in extracting complex information from the input medical images. Apart from the inherent limitations in the design of this study, the results reflect an early promising potential of MLLMs in medical image analysis applications. 
These findings significantly contribute to the body of literature on AI in healthcare, suggesting the need for more sophisticated and context-aware AI tools. Furthermore, this study advances our understanding of AI's role in identifying synthetic images while offering valuable insights for ophthalmology and pulmonary disease diagnosis. It provides important directions for future research and development in this rapidly evolving field. 

\section*{Acknowledgment}
Open Access funding provided by the Qatar National Library.

%
%
%


\end{document}